\documentclass[twocolumn,showpacs,preprintnumbers,amsmath,amssymb]{revtex4}
\usepackage{color}
\usepackage{graphics}
\usepackage{graphicx}
\usepackage{dcolumn}
\usepackage{bm}
\begin{document}
\title{Magnetic phase diagram of cubic perovskites SrMn$_{1-x}$Fe$_{x}$O$_3$} \author{S. Kolesnik, B.
Dabrowski, J. Mais,  D. E. Brown, R. Feng, O. Chmaissem, R. Kruk,}\altaffiliation[Also at
]{Institute of Nuclear Physics, ul. Radzikowskiego 152, Krak\'ow, Poland.} \author{C. W.
Kimball}\affiliation{Department of Physics, Northern Illinois University, DeKalb, IL 60115}
\date{\today}
\begin{abstract}
We combine the results of magnetic and transport measurements with M\"ossbauer spectroscopy and
room-temperature diffraction data to construct the magnetic phase diagram of the new family of
cubic perovskite manganites SrMn$_{1-x}$Fe$_{x}$O$_3$. We have found antiferromagnetic ordering for
lightly and heavily Fe-substituted material, while intermediate substitution leads to spin-glass
behavior. Near the SrMn$_{0.5}$Fe$_{0.5}$O$_3$ composition these two types of ordering are found to
coexist and affect one another. The spin glass behavior may be caused by competing ferro- and
antiferromagnetic interactions among Mn$^{4+}$ and observed Fe$^{3+}$ and Fe$^{5+}$ ions.
 \end{abstract}


\maketitle

\section{INTRODUCTION}
Perovskite manganites, AMnO$_3$, have been studied in great detail during the past several years
because of very interesting magnetic and electronic properties resulting from competing charge,
exchange, and phonon interactions.\cite{Tokura00}  Insulating A-, C-, CE-, and G- type
antiferromagnetic (AFM), metallic ferromagnetic, and charge or orbital ordering properties can be
tuned over a wide range through the choice of size and charge of the A-site cations which control
the degree of structural distortions and the formal valence of Mn.  Recently, increased interest
has focused on the colossal magnetoresistive effect and the destruction of the charge ordering
induced by substitutions on the Mn-site.\cite{Hebert02}

From the point of view of competing interactions, the stoichiometric SrMn$_{1-x}$Fe$_{x}$O$_3$
system is interesting because it should contain Mn$^{+4}$ ($t_{2g}^3$) and Fe$^{+4}$
($t_{2g}^3e_{g}^1$) ions. The G-type AFM ($T_N = 233$~K) and insulating SrMnO$_3$ can be obtained
in a cubic perovskite form through a two-step synthesis procedure,\cite{Negas70} although many
previous studies focused on the hexagonal phase that is stable in air at $T < 1440^{\circ}$C. We
have recently shown that the G-type AFM phase is preserved for single-valent Mn$^{4+}$ in
Sr$_{1-x}$Ca$_{x}$MnO$_3$ in the cubic, tetragonal and orthorhombic crystal
structures.\cite{Chmaissem01}  $T_N$ is suppressed by the bending of the Mn-O-Mn bond angle from
180$^{\circ}$ and by the variance of the average size of the A-site ion via changes in the Sr/Ca
ratio.  The other end member of the series, SrFeO$_3$, is also a cubic perovskite with a helical
AFM structure ($T_N = 134$ K).\cite{Takeda72} The low resistivity ($\sim 10^{-3} \Omega$ cm) and
metallic character when fully oxygenated \cite{MacChesney65,Yamada02} was considered the reason for
the absence of the Jahn-Teller distortion and orbital ordering of Fe$^{4+}$. Deviations from oxygen
stoichiometry in SrFeO$_{3-\delta}$ lead to a formation of several different oxygen-vacancy-ordered
perovskite structures for $\delta = 1/8, 1/4,$ and $1/2$.\cite{Hodges00} The substitution of Co for
Fe yields a SrFe$_{1-x}$Co$_{x}$O$_3$ compound, which is  ferromagnetic for $x \geqslant 0.2$ with
a large negative magnetoresistance for $0 \leqslant x \leqslant 0.7$.\cite{Abbate02} Several
oxygen-deficient SrMn$_{1-x}$Fe$_{x}$O$_{3-\delta}$ compositions have recently been
studied.\cite{Fawcett00}
 The orthorhombically distorted perovskite CaFeO$_3$ compound was shown to undergo
the charge separation to Fe$^{5+}$ ($t_{2g}^3$) and Fe$^{3+}$ ($t_{2g}^3e_{g}^2$).\cite{Takano77}
The highly energetically stable high-spin configuration was invoked as a reason for this behavior.
This charge disproportionation phenomenon can also be observed in
Ca$_{1-x}$Sr$_{x}$FeO$_{3}$,\cite{Takano83} La$_{1-x}$Sr$_{x}$FeO$_{3}$,\cite{Dann94} and
SrMn$_{1-x}$Fe$_{x}$O$_{3-\delta}$.\cite{Fawcett00}

 In this study, we investigate polycrystalline
SrMn$_{1-x}$Fe$_{x}$O$_3$. We have constructed the magnetic phase diagram for fully oxygenated
 samples. We observe an antiferromagnetic order for Fe content $x \leqslant 0.5$ and $x \geqslant
0.9$. For intermediate Fe content $0.3 \leqslant x \leqslant 0.8$ we observe a spin-glass behavior
with features characteristic of ``ideal'' 3D Ising spin glasses. Increasing the Fe content leads to
significant covalency effects, such as a decrease of resistivity and covalent shortening of the
lattice parameter. We also observe Fe$^{3+}$/Fe$^{5+}$ charge disproportionation in stoichiometric
SrMn$_{1-x}$Fe$_{x}$O$_3$.

\section{EXPERIMENTAL DETAILS}
\begin{figure} \resizebox{8.5cm}{!}{\includegraphics{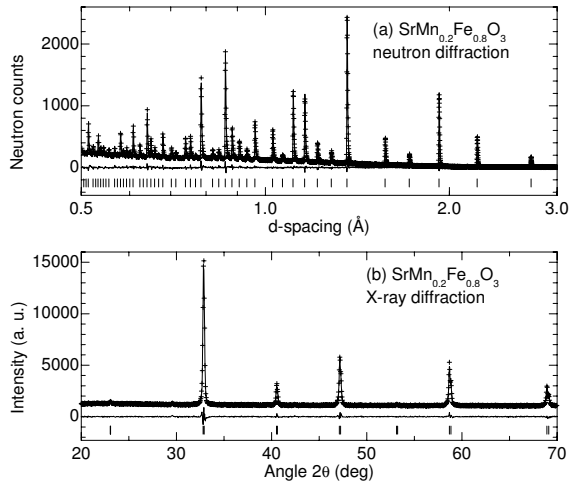}}
\caption{\label{xray} Diffraction patterns for SrMn$_{0.2}$Fe$_{0.8}$O$_3$. Crosses are observed
data points. The solid lines through the data are the Rietveld refinement patterns. The solid lines
below the diffraction patterns represent the differences between the observed and calculated
intensities. The ticks at the bottom mark the peak positions.}
\end{figure}

The samples were prepared using a two-step synthesis method developed for similar kinetically
stable perovskites.\cite{Hinks88} First, oxygen-deficient samples were prepared in argon at $T =
1300 - 1400^{\circ}$C for $x \leqslant 0.5$ and in air at $1300^{\circ}$C for $x > 0.5$. The
samples were then annealed in air or O$_2$ at lower temperatures to achieve stoichiometric
compositions with respect to the oxygen content. High pressure O$_2$ in the range of 140 - 600 bar
was applied for $x \geqslant 0.1$. High-pressure annealing is essential to produce fully oxygenated
samples. The oxygen content in the $x = 0.5$ sample was controlled within the range 2.86 - 3.00 by
annealing the sample under partial pressure of oxygen between $10^{-4}$ and 600 bar on a
thermobalance or in a high-pressure furnace. The samples annealed in the furnace were carefully
weighed before and after annealing and the oxygen content was determined from the mass difference.
The ac susceptibility, dc magnetization and resistivity were measured using a Physical Property
Measurement System Model 6000 (Quantum Design). X-ray diffraction patterns were collected using a
Rigaku diffractometer. Powder neutron diffraction was performed at the Intense Pulsed Neutron
Source at Argonne National Laboratory. Both X-ray and neutron diffraction data were refined using a
GSAS software. Typical diffraction patterns are presented in Fig.~\ref{xray}. M\"ossbauer
measurements were performed in transmission geometry using a 50 mCi $^{57m}$Co in Rh source kept at
room temperature and a krypton proportional detector.  The samples measured at 5 K and 293 K were
placed in an exchange gas cryostat cooled with liquid helium. Silicon diode sensors allowed the
control and stabilization of temperature to within $\pm 0.1$~K.

\section{STRUCTURAL DATA}
\begin{figure}\resizebox{12.5cm}{!}{\includegraphics{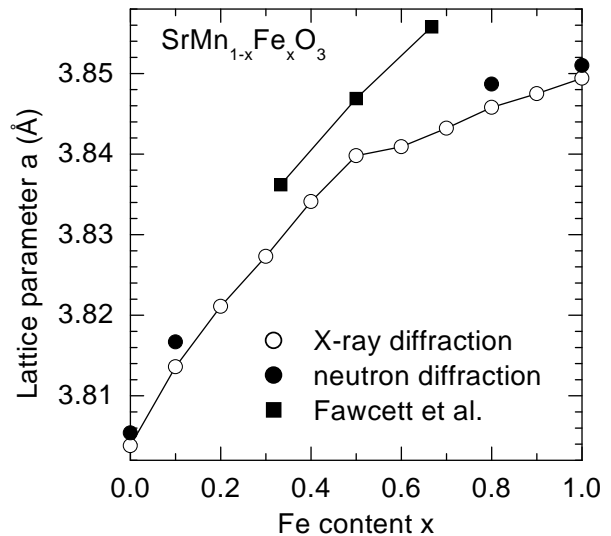}}
\caption{\label{aaxis} Lattice parameter $a$ for SrMn$_{1-x}$Fe$_{x}$O$_3$ samples (circles). Solid
squares are plotted from Ref. \cite{Fawcett00} data. }
\end{figure}

All synthesized samples were single-phase with primitive cubic Pm-3m crystal structure. The
structure can be simply described as a three-dimensional stacking of corner-sharing (Mn,Fe)O$_6$
regular octahedra formed by six equivalent randomly distributed Mn-O or Fe-O bonds.
Fig.~\ref{aaxis} shows the $a$-axis lattice parameter for SrMn$_{1-x}$Fe$_{x}$O$_3$. We also
present previously determined values of $a$ for SrFeO$_3$ \cite{Hodges00}  and
SrMn$_{1-x}$Fe$_{x}$O$_{3-\delta}$ from Ref. \cite{Fawcett00}. The $a$-axis lattice parameter
systematically increases with increasing content of the larger Fe ion substituted for Mn. The slope
of the $a$ vs. $x$ dependence is smaller for larger $x$, which indicates the increasing role of the
covalency of the Fe-O bond.  By studying the structural data for samples with the oxygen content
$3-\delta$ (determined from the thermogravimetric analysis), we also found that the lattice
parameter, $a$, linearly increases with decreasing oxygen content. For example, for $x = 0.5$, the
rate of this increase is 0.064(2) \AA \, per oxygen atom in the formula unit. Hence, we conclude
that the difference of the lattice parameter between our results and those of Ref. 10 is a result
of different oxygen contents.

\section{MAGNETIC PROPERTIES}
\begin{figure} \resizebox{8cm}{!}{\includegraphics{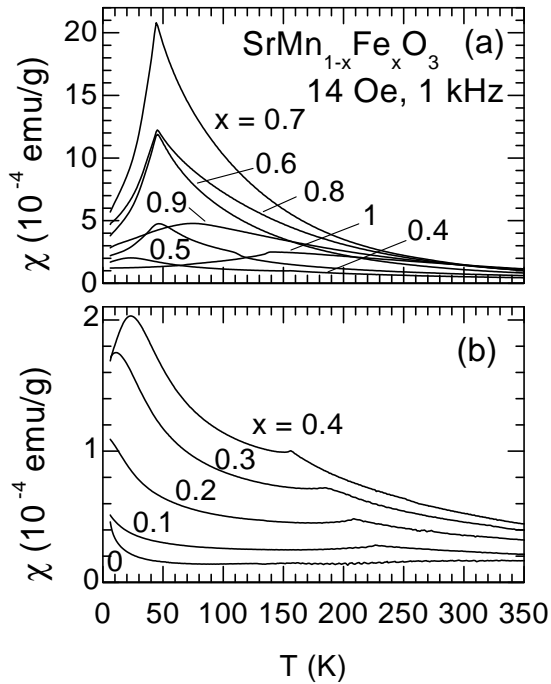}}
 \caption{\label{ac} ac susceptibility for SrMn$_{1-x}$Fe$_{x}$O$_3$ samples. }
\end{figure}

The ac susceptibility for SrMn$_{1-x}$Fe$_{x}$O$_3$ samples is presented in Fig.~\ref{ac}. For $x
\leqslant 0.5$ and $x \geqslant 0.9$, we observe temperature dependencies that are characteristic
of antiferromagnetic materials. For $0.3 \leqslant x \leqslant 0.8$ we also observed a cusp, which
is a signature of spin-glass behavior. From these results we have determined N\'eel temperatures
(defined as the temperatures for which $\chi(T)$ has a maximum slope) and the spin-glass freezing
temperatures, $T_f$ (defined as the temperatures where the susceptibility cusp reaches its
maximum). We also observed additional magnetic properties that substantiate the presence of the
spin-glass state in our samples. These properties will be discussed in detail throughout this
Section. Inverse susceptibility as a function of temperature is linear above $T_N$ for $x < 0.5$
and its intersection with the horizontal axis is negative, which points to antiferromagnetic
interactions. $\chi^{-1}(T)$ for higher Fe contents $x = 0.5 - 1$ is curved and its slope can be
extrapolated either to a negative intersection when we analyze the temperature range just above
$T_N$, or to a positive intersection when we take into account higher temperatures. This behavior
has been observed in SrFeO$_3$ \cite{Takeda72} and is a result of the presence of both
ferromagnetic and antiferromagnetic interactions in these materials.

\begin{figure} \resizebox{12.5cm}{!} {\includegraphics{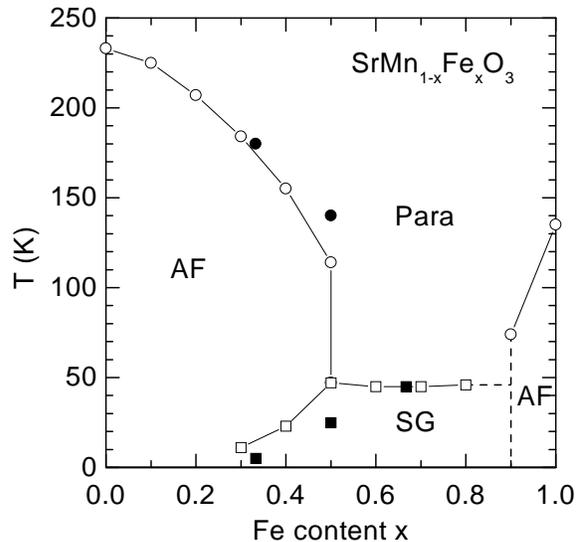} }
 \caption{\label{PhD} Phase diagram for  SrMn$_{1-x}$Fe$_{x}$O$_3$. Solid markers are data points from  Ref. \cite{Fawcett00}.}
\end{figure}

The values of N\'eel temperature and spin-glass freezing temperature are collected in the phase
diagram in Fig.~\ref{PhD}. Four distinct regions in the phase diagram are observed. For $x
\leqslant 0.2$, only an antiferromagnetic phase is observed with $T_N$ decreasing as $x$ increases.
A decrease of $T_N$ has been explained for isoelectronic Ca- and Ba- lightly substituted in the
cubic perovskite SrMnO$_3$ \cite{Chmaissem01} to be a result of the $A$-site size variance
$\sigma^2 = \Sigma y_ir_i^2 - (\Sigma y_ir_i)^2$, where $r_i$ is the ionic size and $y_i$ is the
fractional occupancy of the $A$ site.\cite{RM96} This parameter describes the local variations of
the Mn-O-Mn bond angle in the cubic region that exist even when the average Mn-O-Mn bond angle is
equal to 180 degrees. In the present case, the increase of the Fe content $x$ increases the
$B$-site size variance. This effect changes the local variation of the Mn-O-Mn bond angle even when
the average structure is cubic and hence leads to lower $T_N$. Additionally, different magnetic
$B$-site ions (Fe$^{3+}$ or Fe$^{5+}$: see Sec.~\ref{sec:moss}) randomly substituted for Mn$^{4+}$
change the net exchange integral and introduce disorder, which also lowers $T_N$.

For $x = 0.3 - 0.5$, we observed both antiferromagnetic order and spin-glass behavior. Fawcett {\em
et al.}'s results \cite{Fawcett00} are shown in Fig.~\ref{PhD} for comparison. Fawcett {\em et al.}
also observed both antiferromagnetism and spin glass in the $x = 1/3$ and $x = 1/2$ samples. We
have seen that for a given Fe content, when the oxygen content is increasing, $T_N$ decreases and
$T_f$ increases. Therefore, the $x = 1/2$ oxygen deficient sample shows higher $T_N$ and lower
$T_f$.

In the next region of the phase diagram, where $0.6 \leqslant x \leqslant 0.8$, only the spin-glass
behavior can be observed. The spin-glass freezing temperature is almost constant in this region.
This characteristic temperature is also nearly independent of the oxygen content. The magnitude of
the ac susceptibility (see Fig.~\ref{ac}) is the largest for the $x = 0.7$ sample, which indicates
the largest effective magnetic moment for this composition. The last region is close to $x = 1$,
where only antiferromagnetic order can be observed.

\begin{figure} \resizebox{12.5cm}{!}{\includegraphics{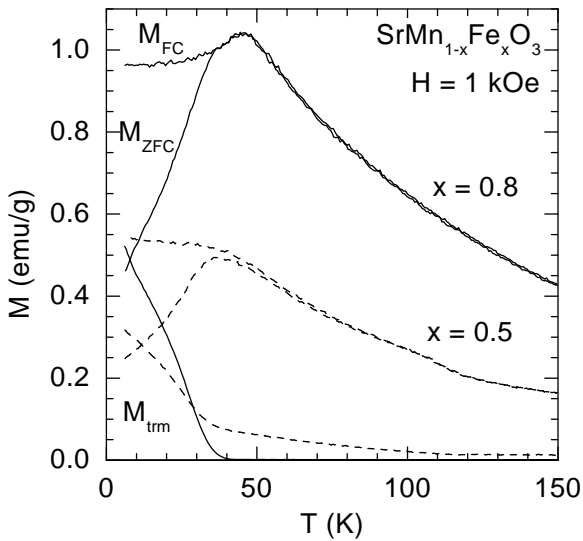}}
 \caption{\label{zfr} ``Zero-field-cooled'' ($M_{ZFC}$), ``field-cooled'' ($M_{FC}$), and
thermoremanent ($M_{trm}$) magnetizations  for  SrMn$_{0.5}$Fe$_{0.5}$O$_3$ (dashed lines) and
SrMn$_{0.2}$Fe$_{0.8}$O$_3$ (solid lines) samples.
 }
\end{figure}

\begin{figure} \resizebox{7cm}{!}{\includegraphics{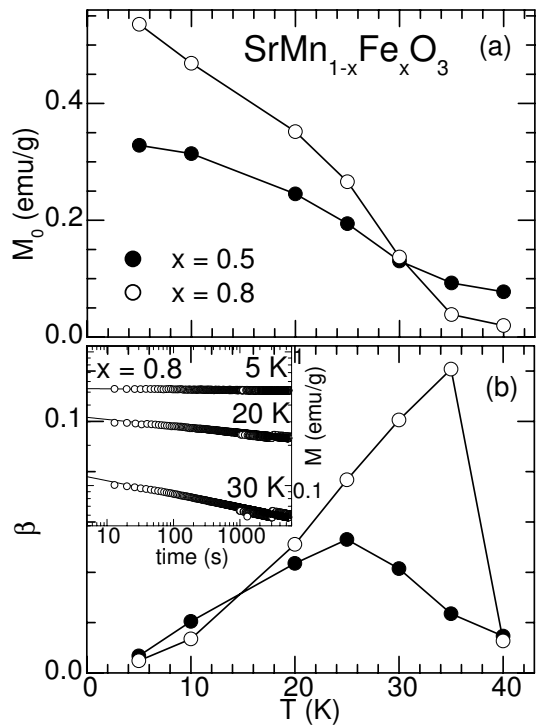}}
  \caption{\label{time} Parameters of time decay of the thermoremanent magnetization for SrMn$_{0.2}$Fe$_{0.8}$O$_3$
 (open circles) and SrMn$_{0.5}$Fe$_{0.5}$O$_3$ (solid circles). The lines are a guide to the eye.
 The parameters $M_0$ (a) and $\beta$ (b) were determined from the fit using $M_{trm} = M_0
 t^{-\beta}$. The inset to panel (b) shows the time dependence of $M_{trm}$ for
 SrMn$_{0.2}$Fe$_{0.8}$O$_3$ and the fits of the above formula to the experimental data.}
\end{figure}

The ``zero-field-cooled'' ($M_{ZFC}$) and ``field-cooled'' ($M_{FC}$) magnetizations, presented in
Fig.~\ref{zfr} for $x = 0.5$ and $x = 0.8$, were measured in the magnetic field of 1 kOe. $M_{ZFC}$
was measured on warming after cooling in a zero magnetic field and switching the magnetic field on
at $T = 5$~K. $M_{FC}$ was subsequently measured on cooling in the magnetic field. We can observe a
difference between $M_{ZFC}$ and $M_{FC}$ below a certain temperature. This difference is typical
for spin glass systems.
 Thermoremanent magnetization ($M_{trm}$), which can be observed after
field cooling to a temperature below $T_f$ and switching off the magnetic field, is also a
manifestation of the spin-glass behavior. The $x = 0.8$ sample shows this difference between
$M_{ZFC}$ and $M_{FC}$ below a certain ``irreversibility temperature''
($T_{irr}$).\cite{deAlmeida78}  $T_{irr} \sim 36$~K and is lower than $T_f$. $M_{trm}$ decreases to
zero at $T_{irr}$ with increasing temperature. This sample shows a transition from the spin glass
state to the paramagnetic state. The $x = 0.5$ sample, which undergoes a transition from spin glass
to the antiferromagnetic state, shows a significant difference between $M_{ZFC}$ and $M_{FC}$ above
$T_f$. In addition, the thermoremanent magnetization can also be observed in the antiferromagnetic
state above $T_f$ up to $T_N$. This observation indicates substantial disorder in the
antiferromagnetic state for the $x = 0.5$ sample. This phase is analogous to the ``random
antiferromagnetic state'' observed in Mn$_{1-x}$Fe$_{x}$TiO$_3$.\cite{Ito88} Thermoremanent
magnetization exhibits a slow decay in time, which is shown in the inset to Fig.~\ref{time} (b).

We have fitted the formula \cite{Ogielski85}
\begin{equation}\label{timedecay}
M_{trm} = M_0  t^{-\beta}
\end{equation}
 to our experimental data and determined the parameters $M_0$ (the extrapolated to zero time magnetization) and
the exponent $\beta$,  which describe the dynamics of spin glasses.  Eq. (\ref{timedecay})
satisfactorily describes the time dependence of the thermoremanent magnetization in nearly the
entire time window we span, except for short times $t < 500$ s where we can see some negative
deviations from this time dependence. This formula, derived by Ogielski in Monte-Carlo simulations,
was used to describe the time decay of three-dimensional Ising spin glasses,\cite{Ogielski85} and
also applied to Mn$_{0.5}$Fe$_{0.5}$TiO$_3$ \cite{Ito86} considered to be an ``ideal''
three-dimensional short-range Ising spin glass.\cite{Mydosh93} The temperature dependence of  $M_0$
and $\beta$ for $x = 0.5$ and $x = 0.8$ samples is presented in Fig.~\ref{time}. $M_0$
systematically decreases with $T$ for both samples. It approaches zero at the irreversibility line
for the $x = 0.8$ sample. For the $x = 0.5$ sample, $M_0$ remains substantially nonzero above $T_f$
in the ``random antiferromagnetic state''. The parameter $\beta$, which is a measure of the
relaxation rate, for $x = 0.8$ increases with $T$ up to $T_{irr} = 36$ K and rapidly drops above
this temperature. The increase of $\beta(T)$ is not exponential as expected for an ``ideal'' spin
glass.\cite{Ito86} The $\beta(T)$ dependence is different for the $x = 0.5$ sample, which shows a
maximum at $T \sim 0.5 T_f$ and a subsequent decrease. This behavior may be a result of the
presence of antiferromagnetic order in the spin-glass state, which inhibits the decay rate of the
thermoremanent magnetization.

\begin{figure} \resizebox{8cm}{!}{\includegraphics{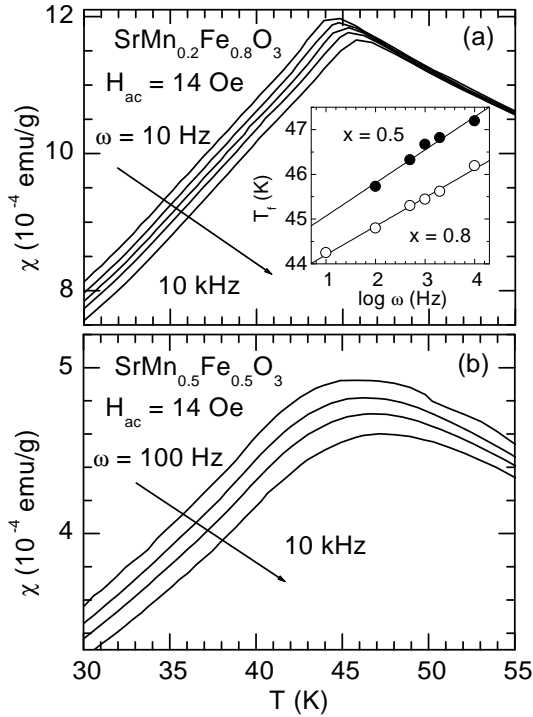}}
 \caption{\label{freq} Temperature dependence of ac susceptibility  for
 SrMn$_{0.2}$Fe$_{0.8}$O$_3$(a) and SrMn$_{0.5}$Fe$_{0.5}$O$_3$ (b)
at several frequencies. Inset shows the linear dependence of $T_f$ on log frequency.}
\end{figure}

 In Fig.~\ref{freq},
we present the ac susceptibility for  SrMn$_{0.2}$Fe$_{0.8}$O$_3$ and SrMn$_{0.5}$Fe$_{0.5}$O$_3$
samples measured at several frequencies in an ac magnetic field of constant amplitude $H_{ac} = 14$
Oe. One can observe a decrease of the ac susceptibility below $T_f$ with increasing frequency, and
a shift of $T_f$ towards higher temperatures. This confirms that the observed cusp in the ac
susceptibility is related to spin-glass behavior.\cite{Mydosh93} For the $x = 0.8$ sample, the ac
susceptibility is independent of frequency above $T_f$ (in the paramagnetic state). For the $x =
0.5$ sample, a significant frequency dependence of $\chi$ can still be observed, which again points
to a frustration of the antiferromagnetic state above the spin-glass freezing temperature. The
inset to Fig.~\ref{freq} (a) shows the dependence of $T_f$ on log frequency. The linear fit to
$T_f(\log\omega)$ gives relative temperature shift vs. frequency $\Delta T_f/[T_f \Delta(\log
\omega)] = 0.0147 \pm 0.008$ and $0.0167 \pm 0.017$ for $x = 0.8$ and $x = 0.5$, respectively.
These values are similar to those observed for canonical spin glasses such as $Pd$Mn and
$Ni$Mn.\cite{Mydosh93}

\section{RESISTIVITY}
\begin{figure} \resizebox{8cm}{!} {\includegraphics{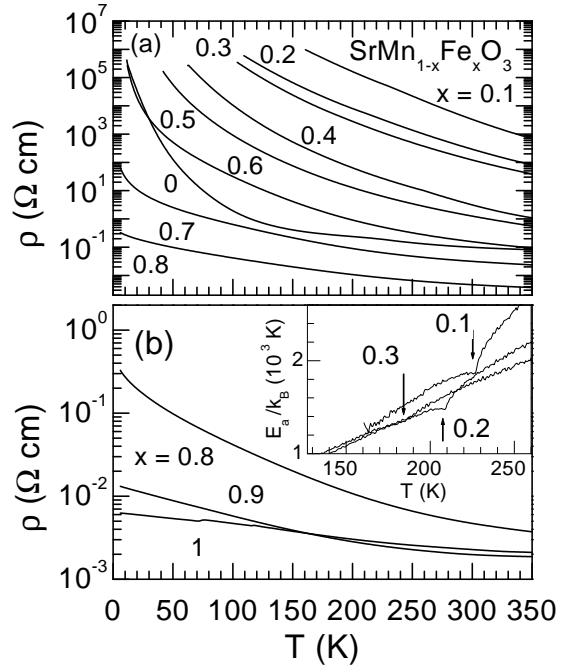} }
 \caption{\label{res} Temperature dependence of resistivity  for  SrMn$_{1-x}$Fe$_{x}$O$_3$ samples. The inset shows the calculated derivative $E_a/k_B
= d\ln(\rho)/d(1/T)$ for $x = 0.1 - 0.3$.}
\end{figure}

The temperature dependence of resistivity  for  SrMn$_{1-x}$Fe$_{x}$O$_3$ samples is presented in
Fig.~\ref{res}. The resistivity is relatively low ($\sim 1$ $\Omega$ cm at room temperature) for
SrMnO$_3$ and increases by over four orders of magnitude on substitution of 10\% Fe for Mn. It
reaches a maximum for $x = 0.1$ and decreases with further Fe substitution. $\rho(T)$ shows mostly
semiconducting dependence.  We were able to fully oxygenate SrFeO$_{3}$ only in the powder form
under high pressure. This powder was used for magnetic, structural, and  M\"ossbauer measurements,
but could not be used for resistivity measurements. Polycrystalline pellets were synthesized, but
were slightly oxygen-deficient. Therefore, these pellets of SrFeO$_{3-\delta}$ show a small
increase of resistivity on decreasing temperature as seen in Fig.~\ref{res}. Assuming thermally
activated resistivity, $\rho = \rho_0exp(E_a/k_BT)$, we estimated the activation energy for
SrMn$_{1-x}$Fe$_{x}$O$_3$. The inset to Fig.~\ref{res}(b) shows the calculated derivative $E_a/k_B
= d\ln(\rho)/d(1/T)$ for $x = 0.1 - 0.3$. We have found that calculated in this way $E_a/k_B$ is
temperature dependent (not constant with respect to temperature as expected for the simple thermal
activation conduction model). The negative kinks, which are marked with arrows, denote $T_N$. We
also checked other models of conduction by introducing a temperature dependent pre-factor to the
formula $\rho \propto T^s exp((T_0/T)^p)$.\cite{Shafarman89} The best description of our data can
be obtained with $s =$ 8 - 9 for this formula. The physical meaning of this value is not yet clear
because this value is much higher than that proposed in the framework of the existing models (e.g.
$s = 1/2$ in Mott's variable range hopping model \cite{Mott69} or $s = 1$ in the small polaron
model \cite{Worledge98}).

\section{M\"OSSBAUER SPECTROSCOPY}
\label{sec:moss} Stoichiometric SrMn$_{1-x}$Fe$_{x}$O$_{3}$ samples were examined by applying
M\"ossbauer spectroscopy on the $^{57}$Fe isotope that is 2\% abundant in the material.  Through a
careful analysis of the magnetic hyperfine field and the isomer shift, M\"ossbauer spectroscopy
provides a way of ascertaining whether iron is in different chemical or crystallographical
environments, as well as its valence state.  High spin Fe$^{3+}$, Fe$^{4+}$, and Fe$^{6+}$ have
room temperature isomer shifts in the range 0.1 to 0.6, -0.2 to 0.2, and -0.8 to -0.9 mm/s,
respectively (relative to $\alpha -$Fe).\cite{Greenwood71}  There are fewer studies of the isomer
shift for Fe$^{5+}$; therefore it is less well understood. The magnetic hyperfine field is
dominated by the Fermi contact interaction which gives rise to about 550 kOe for high-spin
Fe$^{3+}$ having a mean spin of 5/2 for the $3d$ electrons.  Thus, a general rule is that
$\sim$~110 kOe corresponds to $\sim$~1 Bohr magneton (one unpaired electron).  These rules can
substantially change due to covalency effects.  Increasing the covalency between iron and oxygen
tends to produce lower isomer shifts, and to reduce the effective number of unpaired electrons
which leads to lower magnetic hyperfine fields.  In a metallic material polarized conduction
electrons also affect the magnitude of the hyperfine field.

\begin{figure*} \resizebox{16cm}{!} {\includegraphics{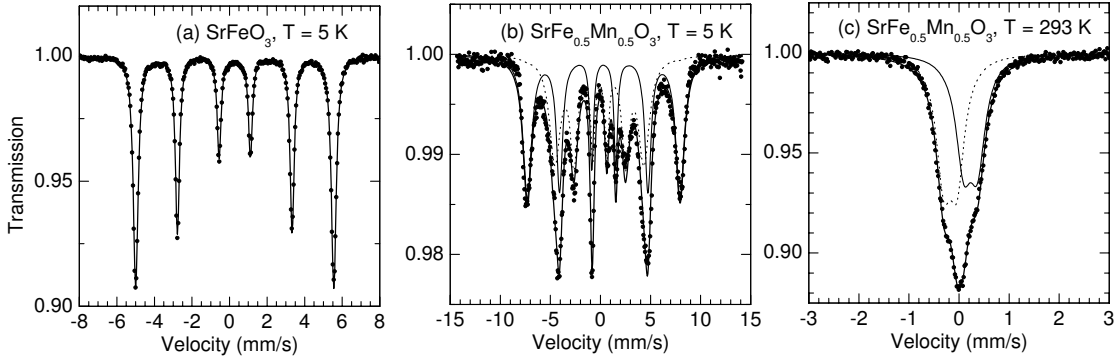} }
\caption{\label{moss} M\"ossbauer spectra  for  SrFeO$_3$ and SrMn$_{0.5}$Fe$_{0.5}$O$_3$ samples.
The fit of the 2-site model for SrMn$_{0.5}$Fe$_{0.5}$O$_3$ is represented by thin solid lines for
Fe$^{3+}$ and dotted lines for Fe$^{5+}$.}
\end{figure*}

The parent compound, SrFeO$_{3}$, is known to be a metallic conductor that orders into a helical
antiferromagnetic structure at low temperatures due to the competition between ferromagnetic
nearest neighbors and antiferromagnetic next nearest neighbors.\cite{Takeda72}  As seen in
Fig.~\ref{moss}, SrFeO$_{3}$ exhibits a set of sharp magnetically split lines (0.24 mm/s linewidth)
at 5~K.  In the paramagnetic state at 293~K, the spectrum consists only of a slightly broadened
single line (0.27~mm/s linewidth). Thus, Fe exists in only one valence state from 5~K to room
temperature. Charge balance suggests that SrFeO$_{3}$ forms with iron in the +4 valence state. This
is confirmed by the measured isomer shift of 0.059 mm/s at 293~K (0.154~mm/s at 5~K). The low
magnetic field of 327 kOe would also indicate that this material has a magnetic moment of $3 \mu_B$
rather than the expected $4 \mu_B$, but the low magnetic field may be due to the nearly delocalized
character of the electron in the $e_g$ orbital of Fe$^{4+}$ with comcomitant lowering of the net
field by the polarized conduction electrons.  The quadrupole splitting for this material is nearly
zero ($\sim -0.019$~mm/sec) at 293~K, agreeing with the values reported in the
literature.\cite{Gallagher64} This result indicates the absence of any extensive static Jahn-Teller
distorted FeO$_6$ octahedra even though high-spin Fe$^{4+}$ is a likely candidate for a Jahn-Teller
ion due to the single electron in the $e_g$ orbital.\cite{Takano77}  However, a dynamic Jahn-Teller
effect is not ruled out.  Due to the lifetime of the excited nuclear state, the $^{57}$Fe nucleus
is sensitive to fluctuating environments that fluctuate on a time scale longer than 10$^{-11}$
seconds. A fast dynamic Jahn-Teller effect where the electronic hopping is enhanced by the
delocalized character of the $e_g$ electron could result in electric quadrupole fluctuation times
that are too short for the Fe nuclei to detect.  Thus, the electric quadrupole interaction
effectively averages to near zero. Thus, the high pressure synthesis technique appears to be
successful in producing highly stoichiometric compounds of metallic  SrFeO$_{3}$ and is a sound
technique to produce stoichiometric samples of SrMn$_{1-x}$Fe$_{x}$O$_{3}$.

Measurements were made on SrMn$_{1-x}$Fe$_{x}$O$_{3}$ at various temperatures to characterize the
charge disproportionation properties of iron in the spin-glass, antiferromagnetic, and paramagnetic
phases.  Charge disproportionation has been observed in the highly nonstoichiometric form of
SrMn$_{1-x}$Fe$_{x}$O$_{3-\delta}$ as well as in CaFeO$_{3}$.\cite{Fawcett00,Takeda78}  Charge
balance would dictate that the valence state of Fe should be +4 (since Fe and Mn share the same
crystallographic sites for the first material and Mn exists in only the +4 valence state
\cite{Fawcett00}).  What is usually observed is a 2-site iron M\"ossbauer spectrum indicating that
Fe exists in 2 valence states rather than in a single, pure valence state of +4. The existence of
two different Fe valence states provides evidence for a charge disproportionation as follows:  2
Fe$^{4+} \rightleftarrows$ Fe$^{(4-\delta)+}$ + Fe$^{(4+\delta)+}$, where $\delta$ varies from 0 to
1 depending on the covalency between iron and oxygen.

We have observed the presence of charge disproportionation in highly stoichiometric forms of
SrMn$_{1-x}$Fe$_{x}$O$_{3}$.  Thus, oxygen vacancies are not a factor in the  charge
disproportionation properties.\cite{Fawcett00}  As Mn was added to SrFeO$_{3}$, the spectra
revealed the presence of two distinct Fe sites having different magnetic hyperfine fields and
isomer shifts. The spectra for one of the samples, SrMn$_{0.5}$Fe$_{0.5}$O$_{3}$, are shown in
Fig.~\ref{moss} and data for it is given in Table I. In this Table, $\varepsilon$  is the effective
\begin{table}\caption{Table I. Hyperfine parameters deduced from M\"ossbauer spectra. $H_{eff}$ is the mean
magnetic hyperfine field; $\varepsilon$  is the effective quadrupole splitting;  $\delta_{is}$ is
the isomer shift relative to $\alpha$-Fe, and $\Gamma_3$ and $\Gamma_1$ are the linewidths of the
inner and outer lines, respectively, in the spectra (at 293 K, there is only a single line), and
Area is the relative area under each subspectra. The areas for the two charge states of Fe in
SrMn$_{0.5}$Fe$_{0.5}$O$_3$ are equal within statistical error.}

\begin{tabular}{|c|c|c|c|c|c|c|}
\hline Compound  &  $T$ & $H_{eff}$ &  $\varepsilon$ & $\delta_{is}$ &$\Gamma_3/ \Gamma_1$ & Area\\
                 &  (K) & (kOe)& (mm/s) & (mm/s) &(mm/s)  & (\%)\\
\hline\hline SrMn$_{0.5}$Fe$_{0.5}$O$_3$  & 5 & 480(1) & -0.009(9) & 0.435(5)  & 0.4/0.99& 52\\
                       &   & 279(1)& -0.04(2)& 0.03(1)& 0.52/1.5 & 48\\
\hline
                       & 293 & -- &  0.289(4)  &   0.350(3)  &  0.39  &   48  \\
                    &         & -- & 0.265(3) & -0.059(3) & 0.36 & 52\\
 \hline   SrFeO$_3$  &           5 &  327(1) & -0.001(1) &  0.154(1) &   0.24  &  100\\
    & 293& -- & -0.02(4) & 0.059(1) & 0.27 &              100\\
    \hline

\end{tabular}
\end{table}
quadrupole splitting, which is defined as $\varepsilon = eQV_{zz} (3*\cos^2\theta-1)/4$ in a
magnetically ordered state and $\varepsilon = eQV_{zz}/2$ in a paramagnetic state where $eQ$ is the
nuclear quadrupole moment, $V_{zz}$ is the z-component of the electric field gradient tensor, and
$\theta$ is the angle between the $V_{zz}$ axis and $H_{eff}$. From this data the Fe ions appear to
exist, in nearly equal proportions, in two valence states in the paramagnetic, antiferromagnetic,
and spin-glass phases (by examining the area under each M\"ossbauer subspectra). Mn is known to
exist in a valence state close to +4 through x-ray absorption spectroscopy chemical shift
measurements on nonstoichiometric compounds.\cite{Fawcett00} Since Mn does not charge
disproportionate, then, because the Fe sites occur in nearly equal proportions, the Fe site having
the largest magnetic field (480 kOe at 5 K) was assigned a +3 valence state, and a +5 valence state
was assigned to the other Fe site.  The Fe site assigned a +3 valence state also has an isomer
shift (0.35 mm/sec at 293 K) that lies near that expected for Fe$^{3+}$ (between 0.1 and 0.6 mm/s
for a high-spin state).  This is exactly what one would expect for fully oxygenated
SrMn$_{0.5}$Fe$_{0.5}$O$_{3}$ -- that the average valence state of Fe would be +4.

\begin{figure} \resizebox{13cm}{!} {\includegraphics{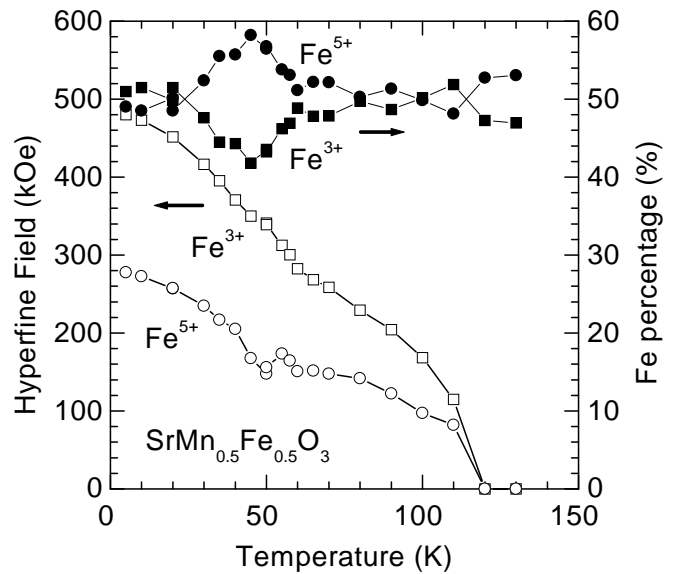} }
\caption{\label{hyper} Hyperfine fields (open markers) and the percentage of Fe$^{3+}$ and
Fe$^{5+}$ (solid markers) determined from M\"ossbauer spectra  for  SrMn$_{0.5}$Fe$_{0.5}$O$_3$.
The spectra were fitted using a 2-site Lorentzian model. }
\end{figure}

The actual environment around each Fe atom is rather complicated. Fe$^{3+}$, Fe$^{5+}$, and
Mn$^{4+}$ can all occupy the same site in this cubic perovskite.  Thus, since each ion has a
different ionic size, there must be a disordered array of Fe$^{3+}$, Fe$^{5+}$, and Mn$^{4+}$
oxygen octahedra of differing sizes and distortions throughout the lattice. It is worth noting that
for fully oxygenated samples the quadrupole splitting at room temperature is smaller than for
nonstoichiometric ones.\cite{Fawcett00}  This reflects a less distorted local environment, without
oxygen vacancies, for the stoichiometric compositions. The distribution of different ions over the
crystallographic lattice gives rise to the slight broadening of the M\"ossbauer lines in the
paramagnetic region.  In addition, there are competing ferromagnetic and antiferromagnetic
interactions in both the spin-glass region and the antiferromagnetic region (since the parent
compound, SrFeO$_{3}$, has a helical antiferromagnetic structure). The much broader M\"ossbauer
lines observed in the magnetically ordered regions for both the antiferromagnetic and spin-glass
phases (the outer linewidths were all greater than 1 mm/sec over the whole temperature region in
Fig.~\ref{hyper}) are likely due to competing magnetic interactions between the transition metal
cations, Fe$^{+3}$, Fe$^{+5}$, and Mn$^{+4}$.  Since the materials under study have a random
distribution of Fe$^{+3}$, Fe$^{+5}$, and Mn$^{+4}$ ions throughout the lattice, different sizes
and magnetic moments of these ions lead to a variation in the structural and magnetic environments
around the Fe sites, resulting in the distribution of hyperfine parameters observed in the
Mossbauer spectra. Thus, as a function of temperature and as a function of composition, from $0.1
\leqslant x \leqslant 0.9$, there is a pronounced broadening of the M\"ossbauer linewidths in both
the antiferromagnetic and spin-glass phases. Fitting the spectra with a distribution of magnetic
hyperfine fields, quadrupole splittings, and isomer shifts did not reveal any clearly
distinguishable characteristics between the antiferromagnetic and spin-glass phases -- there was a
significant amount of frustration in both phases arising from the competing magnetic interactions
that also occur in both phases. However, since the spectra revealed two easily distinguishable
sites, the fits to the spectra were made using a simple Lorentzian model which gives average
hyperfine field values.  The variation of the average magnetic hyperfine field with temperature for
SrMn$_{0.5}$Fe$_{0.5}$O$_{3}$ is given in Fig.~\ref{hyper}.  Due to the complicated nature of the
magnetic interactions, the magnetization curves do not follow the typical $S = 5/2$ or $S = 3/2$
Brillouin curves. The equal proportion of Fe$^{3+}$ to Fe$^{5+}$ is nearly maintained over all
temperatures. However, near the spin-glass/antiferromagnetic transition temperature, there is a
deviation from this behavior indicating the presence of more complex behavior.  Three-site models
have been applied to other materials where Fe charge disproportionates into +3 and +5 valence
states, such as in La$_{1-x}$Sr$_{x}$FeO$_{3}$.\cite{Dann94}  However, applying such a model does
not eliminate the anomaly near the spin-glass transition temperature. Further investigations are
under way to understand the existence of this valence state anomaly near the
spin-glass/antiferromagnetic phase boundary.

\section{SUMMARY}
 In summary, we have studied the cubic perovskite SrMn$_{1-x}$Fe$_{x}$O$_3$ ($0 \leqslant x \leqslant 1$) system.
 By ac susceptibility studies of fully  oxygenated samples, we have constructed the magnetic phase diagram. We
have found antiferromagnetic ordering for the lightly and heavily substituted material, while
intermediate substitution leads to spin-glass behavior. Close to the SrMn$_{0.5}$Fe$_{0.5}$O$_3$
composition these two types of ordering coexist and affect one another. By M\"ossbauer
investigations, we have observed the presence of charge disproportionation of iron to nearly equal
proportion of Fe$^{3+}$ to Fe$^{5+}$ in stoichiometric SrMn$_{0.5}$Fe$_{0.5}$O$_{3}$ and the
single-valence state of Fe$^{4+}$ in SrFeO$_{3}$. The increase of Fe content, $x$, is accompanied
by stronger covalency of the Fe-O bond, which leads to the shortening of this bond,  a decrease of
resistivity, lower isomer shifts and magnetic hyperfine fields.

\section*{ACKNOWLEDGMENTS}
This work was supported by the DARPA/ONR and the State of Illinois under HECA.

\end{document}